# Thermal Conductance across Harmonic-Matched Epitaxial Al-Sapphire Heterointerfaces

Zhe Cheng,[1] Yee Rui Koh,[2] Habib Ahmad,[3] Renjiu Hu,[4] Jingjing Shi,[1] Michael E. Liao,[5] Yekan Wang,[5] Tingyu Bai,[5] Ruiyang Li,[6] Eungkyu Lee,[6] Evan A. Clinton,[3] Christopher M. Matthews,[3] Zachary Engel,[3] Luke Yates,[1] Tengfei Luo,[6,7] Mark S. Goorsky,[5] William Doolittle,[3] Zhiting Tian,[4] Patrick E. Hopkins,[2,8,9] Samuel Graham[1,10,*]

[1] George W. Woodruff School of Mechanical Engineering, Georgia Institute of Technology, Atlanta, Georgia 30332, United States

[2] Department of Mechanical and Aerospace Engineering, University of Virginia, Charlottesville, Virginia 22904, United States

[3] School of Electrical and Computer Engineering, Georgia Institute of Technology, Atlanta, GA, 30332, United States

[4] Sibley School of Mechanical and Aerospace Engineering, Cornell University, Ithaca, New York 14853, United States

[5] Materials Science and Engineering, University of California, Los Angeles, Los Angeles, CA, 91355, United States

[6] Department of Aerospace and Mechanical Engineering, University of Notre Dame, Notre Dame, Indiana 46556, United States

[7] Center for Sustainable Energy of Notre Dame (ND Energy), University of Notre Dame, Notre Dame, Indiana 46556, United States

[8] Department of Materials Science and Engineering, University of Virginia, Charlottesville, Virginia 22904, United States

[9] Department of Physics, University of Virginia, Charlottesville, Virginia 22904, United States

[10] School of Materials Science and Engineering, Georgia Institute of Technology, Atlanta, Georgia 30332, United States

## Abstract

A unified fundamental understanding of interfacial thermal transport is missing due to the complicated nature of interfaces which involves complex factors such as interfacial bonding and mixing, surface chemistry, crystal orientation, roughness, and interfacial disorder. This is especially true for metal-nonmetal interfaces which incorporate multiple fundamental heat transport mechanisms such as elastic and inelastic phonon scattering as well as electron-phonon coupling in the metal and across the interface. All these factors jointly affect thermal boundary conductance (TBC). Therefore, the experimentally measured interfaces may not be the same as the ideally modelled, thus obfuscating conclusions drawn from previous experimental and modeling comparisons. This work provides a systematic study of interfacial thermal conductance across well-controlled and ultraclean epitaxial (111) Al || (0001) sapphire interfaces, known as harmonic-matched interface. The measured high TBC is compared with theoretical models of atomistic Green's function (AGF) and the non-equilibrium Landauer approach, showing that elastic phonon transport dominates the interfacial thermal transport of the Al-sapphire interfaces. By scaling the TBC with the Al heat capacity and sapphire heat capacity with phonon frequency lower than the max Al phonon frequency, a nearly constant transmission coefficient is observed, indicating that the phonons on the Al side limits the Al-sapphire TBC. This confirms that elastic phonon transport dominates interfacial thermal transport of the Al-sapphire interfaces and other mechanisms play negligible roles. Our work enables a quantitative study of TBC to validate theoretical models of thermal transport mechanisms across metal-nonmetal interfaces, and acts as a benchmark when studying how other factors impact TBC.

**KEYWORDS**: Thermal boundary conductance, epitaxy interface, elastic phonon transport, AGF, TDTR, interfacial thermal transport

Thermal transport across interfaces at macroscopic length-scales is described by the interfacial form of Fourier's law: thermal boundary conductance (TBC, $G$) defines a finite temperature drop ($\Delta T$) for a given heat flux ($Q$) across an interface ($G=Q/\Delta T$). This temperature drop was first observed across copper and liquid helium interface at extremely low temperatures by Kapitza in 1941.[1,2] Experimental studies of interfacial thermal transport did not move to higher temperatures until late 1970s when a transient hot-strip method was developed.[3,4] After that, several more measurements on solid-solid interfaces were reported in 1980s and 1990s with the modifications of the hot-strip method and the development of the picosecond transient thermoreflectance.[5-8] When time-domain thermoreflectance (TDTR) was developed, the experimental research on interfacial heat transport started to attract great attention.[9-20] Different factors, such as interfacial bonding[20-24], interfacial mixing[16,25], pressure[12,19], surface chemistry[17,26], crystalline orientation[15,27,28], roughness[18,29,30], interfacial disorder[31-34], were found to affect TBC.

To explain experimental results, theories of interfacial thermal conductance have been developed since 1950s, such as the acoustic mismatch model (AMM) and the diffuse mismatch model (DMM).[5,6,35] More recently, other theoretical tools were used to calculate TBC, for instance, equilibrium/non-equilibrium molecular dynamics (MD)[36-38], interface conductance modal analysis (ICMA),[39,40] wave packet method[41,42], atomistic Green's function (AGF)[43,44], and non-equilibrium Landauer approach[45]. Due to the complicated nature of interfaces, most of the theoretical calculations cannot capture the detailed features of the interface. It still remains an open question whether the modelled interfaces are the same as the measured ones in most of previous works due to the lack of detailed simultaneous material and thermal characterization of the interfaces.[46] This is especially true for metal-nonmetal interfaces because of the complicated interfacial structures and multiple carrier transport mechanisms near the interfaces. Metal growth on nonmetal substrates, usually by sputtering or evaporation, suffers from problems like chemical reaction or interfacial mixing with substrates during deposition processes, the inclusion of an oxide or contaminating layers at the interface, polycrystalline metal films with a mixture of different orientations, or poor adhesion with substrates.[16,25,32,47-49] These result in the growth of metal-nonmetal interfaces to deviate from the ideal interface often assumed in theoretical modelling, so direct comparison between theoretical and experimental results limits our ability to draw accurate conclusions.[34,50]

Additionally, interfacial thermal transport involves multiple fundamental transport mechanisms, including elastic and inelastic phonon transport across the interface, and electron-phonon coupling in the metal and across the interface.[51-55] The relative contributions of these mechanisms to TBC remain unclear and are still an open question.[51,56] For instance, theoretical calculations show electrons on the metal side can pass some energy directly to phonons on the nonmetal side while experiments show that a 400-fold change in electronic density for otherwise similar metals on the metal side does not impact TBC significantly.[7,13,55,56] Even though some efforts have been made to study epitaxy interfaces,[9,46,57] a clear fundamental understanding of interfacial thermal transport across metal-nonmetal interfaces is still lacking possibly due to the lack of systematic benchmark studies of well-controlled interface growth, simultaneous structural and thermal characterizations, and corresponding comparison with thermal models because very few computational methods can take interface non-idealities into consideration.[44,58,59]

In this work, we fill the gap by epitaxially growing (111) Al on (0001) ultraclean sapphire substrates by Molecular Beam Epitaxy (MBE). The Al-sapphire system is particularly suitable for benchmarking because of the atomically smooth surfaces, easy cleaning by baking at high temperatures in ultrahigh vacuum (UHV) conditions, no surface oxidation during baking, and no reaction with Al during growth.[56] These orientations are selected because of the relatively small lattice mismatch (4%) and similar crystalline structure, as shown in Figure S1. The TBC of Al-sapphire interfaces are measured by three TDTR systems in Georgia Institute of Technology, University of Virginia, and University of Notre Dame. Simultaneous structural characterizations are performed with a transmission electron microscope (TEM) and x-ray diffraction (XRD). The calculated TBC by AGF and non-equilibrium Landauer approaches are compared with experimental data over a temperature range of 80-480 K. The phonon and electron transport mechanisms across interfaces and their contributions to TBC are discussed. Our work provides a benchmark for interfacial thermal conductance across Al-sapphire interfaces and enable quantitative study of TBC to validate theoretical models.

Two samples (Sub100 and Sub200) were studied in the round robin analysis in this work. The substrate temperature for Al growth was kept at 373 K for sample Sub100 and at 473 K for Sub200 while all the other growth conditions remained the same. More details about MBE sample growth can be found in the Methods section. A key step is that the sapphire was cleaned through a high temperature annealing step in UHV prior to the epitaxial deposition of the Al in-situ by MBE. Figure 1(a) shows the XRD patterns of samples Sub100 (the peaks observed for Sub200 are the same as observed for Sub100). The results show that the Al films are effectively single crystalline with the presence of twins. A standard 2θ:ω scan shows only reflections from Al (111) planes

[(111) and (222) reflections)] and the sapphire (0001) planes [(0006) and (000 12) reflections]. As shown in Figure 1(b), Phi scans of the Al (220) and the sapphire (११२̄3) demonstrate that the epitaxial alignment is Al [११२̄] ∥ Al$_2$O$_3$ [११२̄0] which fits with ± Al [१̄10] ∥ Al$_2$O$_3$ [१०१̄0]. Twinning is observed through the phi scans (there are peaks every 60°), showing that the layers include a Σ3 twin-related boundaries (180° rotation around the surface (111)). The peak width (FWHM) for the sapphire (११२̄3) reflections is less than 0.1°. The Al (220) reflection FWHM is about 0.43°; while this is larger than the substrate peak widths, but still small enough to preclude high angle grain boundaries. Figure 1(c) shows an AFM image of the smooth Al surface and the RMS roughness for a 5 µm x 5 µm area are 0.125 nm for Sub200. Figure 1(d) shows a cross-section high-resolution TEM (HRTEM) image of the Al-sapphire interface. The orientations of the Al and sapphire are cross-plane [111] Al ∥ [0001] Al$_2$O$_3$ and in-plane [110] Al ∥ [१०१̄0] Al$_2$O$_3$. We see a sharp, distinct, and well-matched Al-sapphire boundary with evidence of only a sub-nm interfacial re-arrangement.[60] Electron back-scattered diffraction measurements (EBSD) (in SI) confirm the film is completely (111) oriented with the Σ3 twin boundaries and the non-twinned regions extend as much as several tens of microns. More characterization in the SI also shows that the Al films have no strain. We note that these dimensions are more indicative of a high-quality crystalline film than those had been previously obtained for low temperature Al grown on sapphire.[61] These characteristics support the contention that the Al-sapphire interface is an ideal interface which could be compared with modeling results directly.

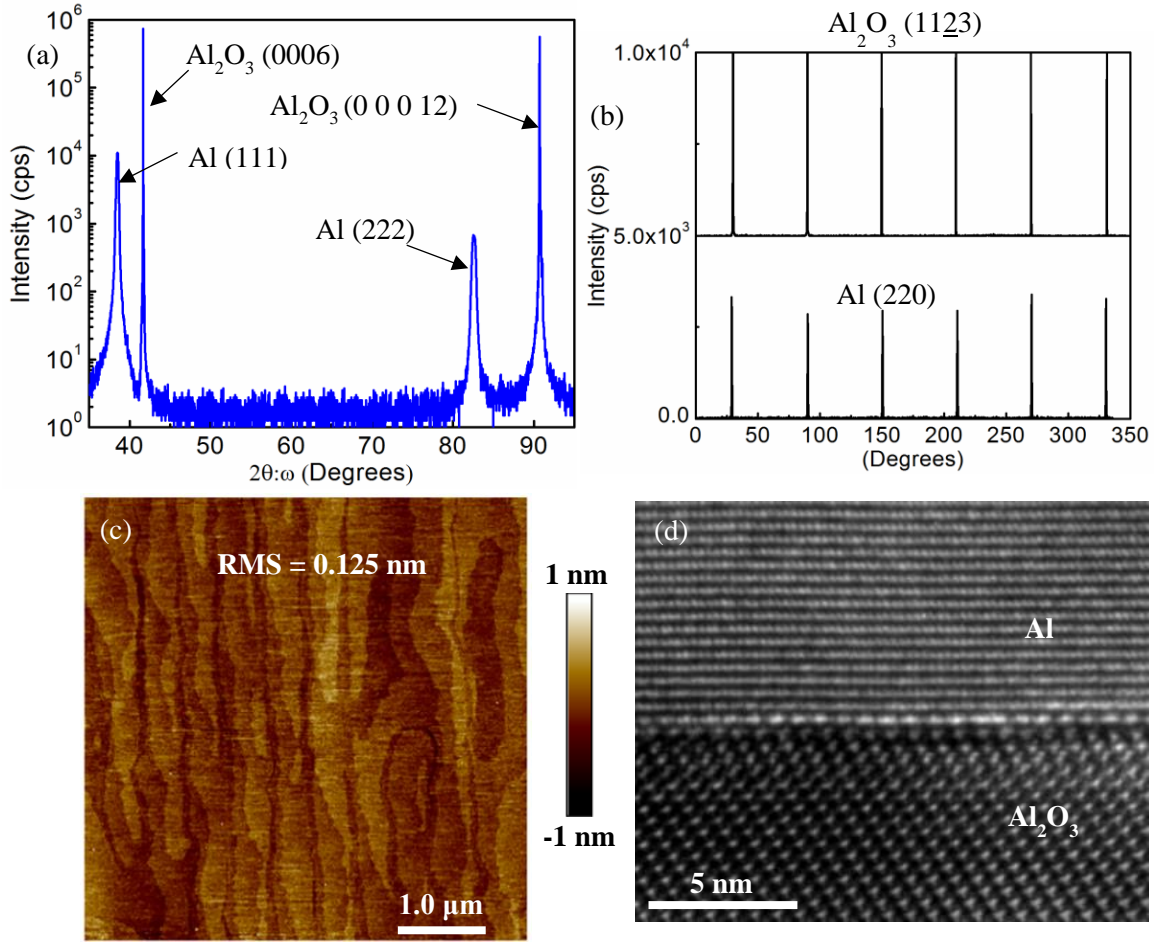

**Figure 1**. (a) XRD patterns of Sub100 on a log scale. (b) Phi scans (220) Al and (११२३) Al$_2$O$_3$. (c) AFM images of Al surface. RMS roughness for a 5 µm x 5 µm area are 0.125 nm for Sub200. (d) HRTEM cross-section image of Al-sapphire interface: cross-plane [111] Al || [0001] Al$_2$O$_3$ and in-plane [110] Al || [10$\bar{1}$0] Al$_2$O$_3$. We see a sharp and distinct Al-sapphire interface.

Figure 2(a) shows the phonon density of state (DOS) of Al and sapphire calculated with density-functional theory (DFT). The maximum phonon frequency of sapphire (about 25 THz) is much higher than that of Al (about 10 THz). If only considering elastic phonon scattering, phonons with frequencies between 10 THz and 25 THz in sapphire do not contribute to TBC. The thermal resistance circuit of thermal transport across Al-sapphire interface is shown in Figure 2(b).

Electrons dominate thermal transport in metals while phonons dominate in non-metals. For thermal transport across metal and non-metal interfaces, thermal energy carried by electrons in the metals needs to be transferred to phonons in metals first. Then phonons in the metals transfer energy across the interface to phonons in non-metals through elastic and inelastic processes. Near the Al-sapphire interface, we assume that the boundary condition for electron transport is adiabatic while phonons in the Al side can transmit through the interface to the sapphire side.[12,13,54] There is a temperature difference between the near-interface electrons and phonons in the Al side. This local non-equilibrium results in a corresponding electron-phonon coupling thermal resistance in Al. Phonons in the Al side transfer energy to phonons in the sapphire side through elastic and inelastic channels. Additionally, some theoretical calculations show electrons in the metal side could directly pass energy to phonons in the nonmetal side while some other calculations and experimental data show that this cross-interface electron-phonon coupling does not contribute to TBC significantly.[7,13,51,55,62,63] We add this possible heat transfer channel in Figure 2(b) as well.

The measured TBC of Al-sapphire interfaces are shown in Figure 2(c) and literature values are included as comparison. The TBC values were measured and consistent results were obtained among different university groups in the round robin. The measured TBC of Sub100 is the same as that of Sub200, indicating the substrate temperature during growth does not affect either the thermal properties or the structure of Al-sapphire interfaces. The measured TBC in this work is larger than all the other Al-sapphire TBC values in the literature.[7,15,64] This may be a result of the impact of the interface non-idealities of Al-sapphire grown by other methods on TBC. This may also give us new insight into the temperature dependence of the TBC of Al-sapphire interfaces and requires additional analysis of imperfect interfaces that can exist when grown by other methods.

For instance, the measured TBC reaches a plateau above room temperature, which has a different trend from the measured values in the literature[64] in which the increased TBC at high temperatures with temperature was attributed to inelastic phonon scattering. Many factors, such as crystalline orientation, roughness, and interfacial disorder or contamination, jointly affect TBC of Al-sapphire interfaces. This highlights the importance of a benchmarking study on TBC of ideal interfaces, which can not only be used to validate theoretical thermal models across perfect interfaces, but also act as a reference when studying how other factors impact TBC.

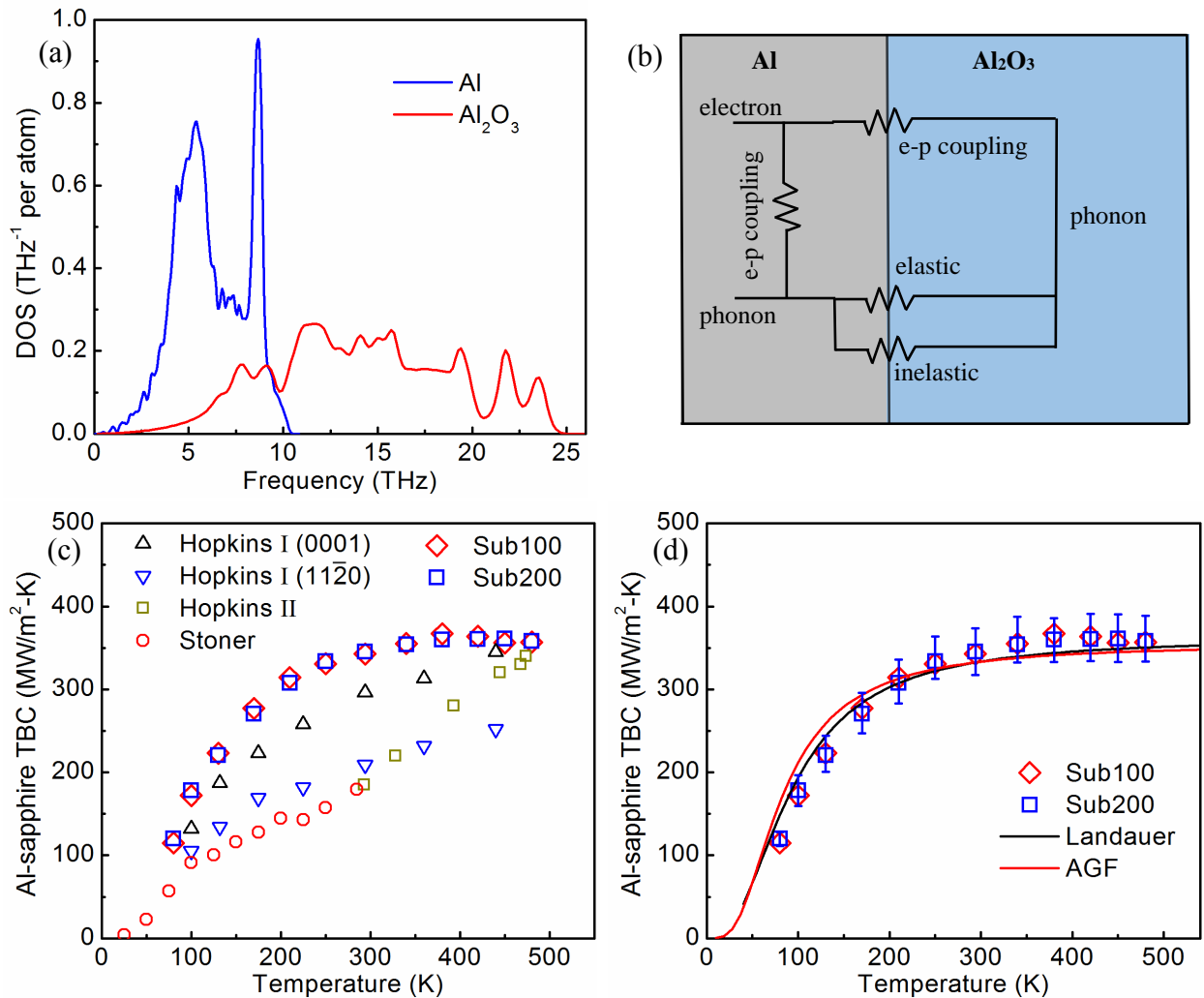

**Figure 2**. (a) phonon DOS of Al and sapphire. (b) thermal circuit of Al-sapphire interfaces. (c) comparison of temperature dependent Al-sapphire TBC of samples Sub100 and Sub200 with measured TBC values in the literature: Hopkins I[15], Stoner[7], and Hopkins II[64]. (d) comparison of measured Al-sapphire TBC with theoretical values calculated by AGF and non-equilibrium Landauer approach.

As shown in Figure 2(d), thermal modeling results are compared with experimental TBC values. Intrinsic Landauer TBC is calculated with a non-equilibrium Landauer approach where typical Landauer formula with DMM are used with further temperature corrections which address the non-equilibrium effect of phonon transport near the interfaces. Intrinsic Landauer TBC assumes incident phonons have the same modal equivalent equilibrium temperature. More details about the non-equilibrium Landauer approach can be found in the Methods section and Ref.[45] The small difference in AGF results between our result and Ref.[65] should be attributed to different force constants. Here we used DFT calculation to generate force constants while Ref.[65] generated the force constants via empirical lattice dynamics calculations. Both non-equilibrium Landauer approach and AGF account for quantum effect of phonon transport at the interfaces while they do not include inelastic contribution to TBC in the calculation. At low temperatures, TBC increases with temperature because of the increasing number of phonons involved interfacial thermal transport. When temperatures go above the Debye temperature of Al (428 K), the TBC reaches a plateau because the phonons in the Al side is fully excited.[66]

As shown in Figure 2(b), several thermal transport mechanisms are involved in interfacial heat transport. The good agreement between experimental data and theoretical values calculated by AGF and non-equilibrium Landauer approaches suggests that elastic phonon transport across the interface is dominant. All the other thermal transport mechanisms are negligible. AGF and non-equilibrium Landauer approach only consider elastic process. The harmonic lattice matched Al-sapphire interface are expected to have weak inelastic process. The electron-phonon coupling thermal resistance in Al is not large as well due to the large electron-phonon coupling constant of Al. We will discuss the role of each heat transfer mechanism below.

First, electron-phonon coupling in the Al side could result in a thermal resistance in series with the interfacial phonon-phonon thermal resistance.[54] The thermal resistance can be estimated as $1/\sqrt{\kappa_p G_{ep}}$.[54] Here, $\kappa_p$ and $G_{ep}$ are the lattice thermal conductivity and electron-phonon coupling constant of Al. We consider $G_{ep}$ independent of temperature and its value at room temperature is 5.38 ×10$^{17}$ W/m$^3$-K.[66] $\kappa_p$ of Al at room temperature is 6 W/m-K based on first-principle calculations.[66] The thermal resistance derived from electron-phonon coupling in Al is 0.557 m$^2$-K/GW, 19% of the measured overall thermal resistance across Al-sapphire interfaces. As the temperature decreases, the lattice thermal conductivity increases, leading to a reduced electron-phonon coupling thermal resistance. The phonon-phonon TBC decreases with temperature, leading to a larger phonon-phonon thermal resistance. As a result, the effect of the electron-phonon coupling thermal resistance on overall TBC would become smaller at low temperatures. However, for temperatures comparable or higher than the Debye temperature of Al (428 K), the lattice thermal conductivity Al decreases with temperature ($\kappa_p \sim 1/T$) so electron-phonon coupling thermal resistance is proportional to $\sqrt{T}$. From 300 K to 480 K considered in this work, the electron-phonon coupling thermal resistance in the metal side could account for about 20% of the overall measured thermal resistance across Al-sapphire interfaces. But we did not observe clear TBC change due to the effect of this mechanism in our experimental results.

In terms of cross-interface electron-phonon coupling for Al-sapphire interface, we tend to believe that it does not contribute to TBC significantly because the measured TBC is so close to the elastic phonon-phonon TBC for the whole temperature range, especially for low temperatures where inelastic phonon contribution and electron-phonon coupling thermal resistance in the metal are not important. This cross-interface electron-phonon coupling is an additional thermal channel across the interface which could increase TBC. However, the modeled TBC by AGF and non-equilibrium Landauer approach are slightly larger than the measured TBC at low temperatures. Furthermore, a recent theoretical work shows that this cross-interface coupling effect for Si-Cu interface contribute slightly to the overall TBC.[63] Other previous experimental measurements also show that cross-interface electron-phonon coupling is not important for interfacial thermal conductance.[7,13]

AGF TBC and Intrinsic Landauer TBC match quite well with measured TBC which shows that elastic phonon contribution to TBC dominates interfacial thermal transport, and inelastic phonon contribution to TBC is negligible for Al-sapphire interfaces. At high temperatures where inelastic phonon contribution to TBC is expected to be larger, the measured TBC keeps constant. As shown in Figure 2(b), electron-phonon coupling in Al adds additional thermal resistance and reduces the overall TBC while inelastic phonon contribution provides an additional thermal channel which increases TBC. Therefore, the inelastic phonon contribution to TBC may counteract the effect of electron-phonon coupling thermal resistance in Al at high temperatures and both do not contribute much to the overall TBC. This contradicts the case of metal-diamond interfaces where inelastic contribution to TBC is so large that the measured TBC is greatly higher than the radiation limit (the maximum TBC only considering elastic contribution).[13,53] It may result from the high phonon DOS mismatch of metal and diamond. But it should be noted that the conclusion about metal-diamond interfaces needs to be revisited because of the unknown interfacial structure of these interfaces.

For a single phonon mode traveling across an interface, the heat flux equals the phonon energy times the phonon velocity and transmission coefficient. If the temperature difference across the interface is small, TBC contributed by this phonon mode is the product of its heat capacity, phonon velocity, and transmission coefficient. Here, an average value of the product of phonon velocity ($v$) and transmission coefficient ($\tau$) of all the phonon modes can be obtained by dividing TBC with volumetric heat capacity ($C_v$), namely, $\langle v\tau \rangle/3 = TBC/C_v$. Here, the factor of 3 in the formula is derived from three directions and $\langle\ \rangle$ means the average over all the phonon modes. As shown in Figure 3(a), very surprisingly, $\langle v\tau \rangle/3$ is almost a constant over the whole temperature range except the two lowest temperature points when we scale TBC with the Al heat capacity and sapphire heat capacity with phonon frequency lower than the max phonon frequency of Al (~10THz). Phonon velocity is very weakly dependent on temperature. As a result, the average transmission coefficient is almost a constant for different temperatures. This validates theoretical modeling methods such as the AGF and the non-equilibrium Landauer approaches where transmission coefficient is considered to be independent of temperature. This strongly supports that TBC is determined by the number of phonons on the Al side and phonon contribution dominates interfacial thermal transport across Al-sapphire interfaces. The heat-capacity-like TBC hints that phonons with frequency higher than the Al max phonon frequency (10-25 THz) does not contribute much to TBC. The high frequency phonons cannot transmit across interface through elastic processes and cannot contribute to TBC, which confirms that inelastic phonon contribution to TBC is negligible in interfacial thermal conductance of Al-sapphire interfaces.

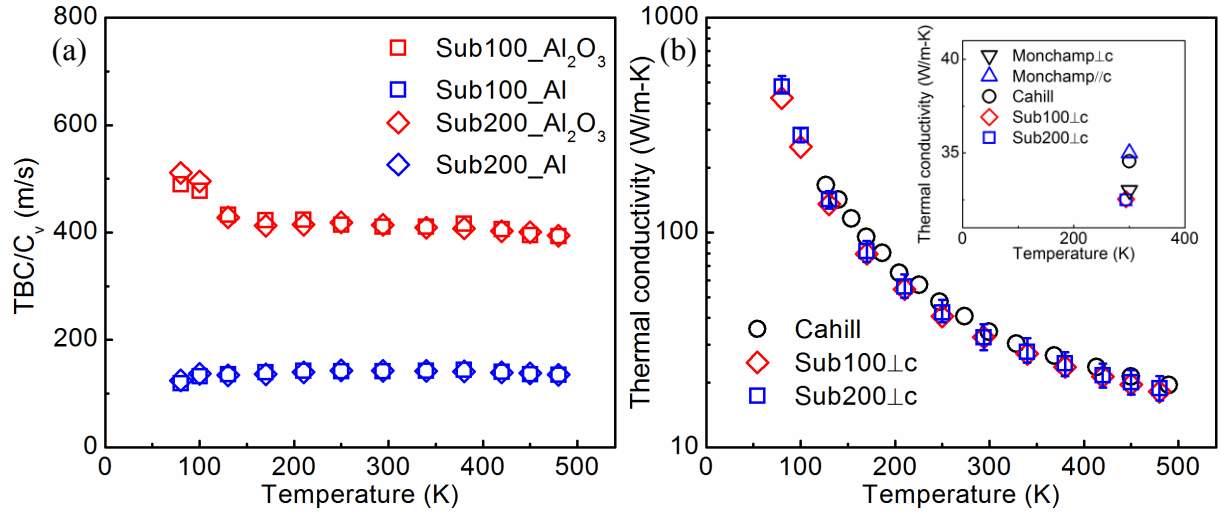

**Figure 3**. (a) TBC scaled with volumetric heat capacity of Al and sapphire with phonon frequency < 10 THz for samples Sub100 and Sub200. (b) temperature dependence of measured sapphire thermal conductivity. The "Cahill" data is from Ref.[67] and the "Monchamp" data is from Ref.[68]

The temperature dependence of the measured sapphire thermal conductivity is shown in Figure 3(b). For Sub100 and Sub200, TDTR measurements are more sensitive to cross-plane direction so the measured thermal conductivity is perpendicular to the c-plane. The measured room temperature value is 32.5 W/m-K, very close to the room temperature value (33 W/m-K) in the Ref.[68] as shown in the inset of Figure 3(b). The thermal conductivity of sapphire is weakly anisotropic for parallel or perpendicular to the c-plane. Sapphire with a crystal orientation parallel to the c-plane has a slightly larger thermal conductivity (35 W/m-K at room temperature). Ref.[67] did not include crystal orientation information in the paper but its measured value is very close to the measured thermal conductivity of sapphire with a crystal orientation parallel to the c-plane in Ref.[68], as shown in the inset of Figure 3(b). Therefore, we speculate that the crystal orientation is parallel to the c-plane. Our temperature-dependent measured sapphire thermal conductivity matches well with the values in Ref.[67] and we attribute the small difference to the weak anisotropy.

In summary, this work provided a systematical benchmark study of interfacial thermal conductance across well-controlled epitaxial (111) Al ∥ (0001) sapphire interfaces with simultaneous structural and thermal characterizations. The measured high TBC is compared than the other Al-sapphire TBC reported in the literature. The analysis of the result presented in this work relied greatly on the concurrent materials and thermal characterization. By having ultraclean epitaxially grown Al-sapphire interfaces, we determined that atomistic Green's function (AGF) and a non-equilibrium Landauer approaches were sufficient for predicting the TBC of Al-sapphire interfaces. The results show that elastic phonon scattering dominates interfacial thermal conductance of Al-sapphire interfaces. Inelastic phonon scattering, electron-phonon coupling within the metal and across interface play negligible roles. The estimated electron-phonon coupling thermal resistance within Al is about 20% of the overall thermal resistance but we did not observe clear TBC change due to this effect. The temperature independence of Al-sapphire TBC at high temperatures suggests that the effect of inelastic phonon scattering is negligible, especially at high temperatures. By scaling TBC with the Al heat capacity and sapphire heat capacity with phonon frequency lower than the Al max phonon frequency, a nearly constant transmission coefficient is observed, indicating the number of phonons in the Al side determines the Al-sapphire TBC. This nearly constant transmission coefficient validates the assumptions in AGF and non-equilibrium Landauer calculations. This demonstrates a method for characterizing the interfacial thermal conductance across metal-nonmetal interfaces. It has enabled a quantitative study of TBC to validate theoretical models, and acts as a benchmark when studying how other quantities impact TBC.

## Methods Section

**Sample Preparation**: In this study, ~80 nm Al was deposited on high temperature annealed sapphire substrates in a Riber 32 Molecular Beam Epitaxy (MBE) system. A multi-step annealing approach was used in order to achieve highly ordered terrace-and-step structure over the surface of sapphire substrates. High temperature annealing reduces surface energy of the sapphire substrates and forms a terrace-and-step structure which acts as nucleation sites for subsequent deposition of Al. 2-inch diameter sapphire wafers were introduced into the front zone of Minibrute furnace and heated under nitrogen environment at 1273 K for 5 minutes. Subsequently, the wafers were moved into the center zone and heated under nitrogen environment at 1384 K for 15 minutes. The sapphire wafers were then annealed at 1384 K for 1 hour under ultra-zero grade air environment followed by 5 hours of heating under the same environment at 1448 K. Finally, the sapphire wafers were cooled down naturally under nitrogen. Tantalum was deposited on the backside of annealed sapphire wafers using a Unifilm DC sputterer to facilitate uniform heating during MBE growth. The backside metallized wafers were diced into 1 cm x 1 cm substrates. These high temperature annealed sapphire substrates were cleaned in a piranha solution of sulfuric and hydrogen peroxide (3:1) at 423 K for 10 to 20 minutes. These substrates were then loaded into a load lock chamber of the MBE and were thermally cleaned under vacuum at 473 K for 10 minutes. The substrates were then transferred into main chamber of the MBE and outgassed at 1073 K for 10 minutes. 80 nm of Al was then deposited on the sapphire substrates from a conventional Veeco effusion cell. Streaky Reflection High Energy Electron Diffraction (RHEED) patterns were observed which implies high surface smoothness. The substrate temperature for Al deposition was kept at 373 K for sample Sub100 and at 473 K for Sub200. Atomic Force

Microscopy (AFM) characterization showed highly smooth Al layers. The top Al layer RMS surface roughness for Sub200 was 0.125 nm and 0.288 nm for Sub100 for 5 µm x 5 µm images.

**TEM:** Plan view and cross-section TEM samples have been prepared with a FEI Nova 600 dual-beam focused ion beam system to characterize the Al layer and sapphire substrate. HRTEM cross-section images were taken at the Al/sapphire interface with a FEI Titan S/TEM operating at 300 keV.

**XRD**: A Bruker JV D1 high-resolution X-ray diffractometer using Cu kα radiation with incident beam conditioning that includes an incident beam parallel beam optical element was employed for pole figures and Φ-scans. The (220) Al reflection Φ scan was measured along with the sapphire (11$\bar{2}$3) reflection. ω:2Θ scans used an additional incident beam monochromating optic (Si (220) two reflection channel cut crystal) of the deposited film on the substrate and were taken to determine which Al orientations were present. The Al lattice parameter was also determined using a high resolution mode with both incident and scattered Si (220) channel cut crystal optics. The Φ scans and the ω:2Θ scans were taken using 0.003 – 0.05° step sizes.

**EBSD**: EBSD images were produced using an FEI Quanta 3D FEG Dual Beam (SEM/FIB) system with an HKL EBSD attachment. The data was obtained with HKL fast acquisition software. Typical step sizes were ~ 2 µm over areas of 250 µm x 250 µm.

**TDTR**: TDTR is an optical pump-probe method for thermal characterization of both bulk and nanostructured materials.[10] A layer of Al is usually coated on the sample surface as transducer.

The thickness of Al is measured by picosecond acoustic technique (83 nm for Sub100 and 86 nm for Sub200). A modulated pump beam heats the sample surface while a delayed probe beam detects the temperature variation of the sample surface via thermoreflectance. The signals picked up by a photodetector and a lock-in amplifier are fitted with an analytical heat conduction solution to infer unknown parameters. The modulation frequency and objective used in this work are 3.6 MHz and 10X objective with pump and probe beam diameters of 19.0 μm and 13.3 μm, respectively. More detail about TDTR can be found in literature.[69,70] The error bars of TDTR measurements are calculated with a Monte Carlo method by considering all the possible error sources.[71] The error bars of beam size measurements by a beam profiler are ±0.5 μm. The error bars of Al and sapphire heat capacity are ±2%. The error bars of Al thickness and Al thermal conductivity are ±3.5% and ±10%, respectively.

**Non-equilibrium Landauer Approach**: The Landauer approach is a widely used method to predict TBC, and the general form of the Landauer formula is from the particle description of phonons and the TBC is calculated from net heat flux and temperature drop across the interface. In previous studies of non-equilibrium effect at the interface, it is pointed out that the phonons are in strong non-equilibrium because of the difference in modal transmission coefficients and reservoir temperatures, and this non-equilibrium effect should be considered in Landauer formula. With the recently developed non-equilibrium Landauer approach which can capture the non-equilibrium effect, the theoretical predictions agree much better with experimental results. We have applied the non-equilibrium Landauer approach to our aluminum-sapphire interface to predict the TBC. In our calculation, the phonon properties of both Al and $Al_2O_3$ are obtained from ab initio calculations within the framework of DFT, as implemented in the Vienna Ab initio Simulation Package (VASP), and the second order force constants are obtained from Phonopy[72].

**AGF**: AGF is a widely used method to calculate the transmission and related thermal properties of a system. More detailed introduction could be found in literature.[43,73,74] With the harmonic assumption, only the second order force constants are needed for Green's function calculation. Here, the second-order DFT force constants of the leads and the interface are separately obtained from the frozen-phonon method[75,76] using QUANTUM ESPRESSO[77] and Phonopy[72]. We used projector augmented-wave method[78] of Perdew, Burke and Ernzerhof[79]. The cut-off energy of the interface is 100 Ryd and the $k$-points mesh is $4 \times 4 \times 2$ for the $2 \times 2 \times 1$ supercell. For the aluminum lead, the cut-off energy is 50 Ryd, and the $k$-points mesh is $4 \times 4 \times 3$ for the $3 \times 3 \times 3$ supercell. While for the sapphire lead, the cut-off energy is 80 Ryd, and the $k$-points mesh is $4 \times 4 \times 2$ for the $3 \times 3 \times 2$ supercell. To represent the infinitely large transverse direction, the transverse $k$-points mesh is $20 \times 20$ in the Brillouin zone to ensure the convergence. In the experiment, Al is oriented to [111] direction to match the lattice constants of sapphire. To calculate the thermal conductivity along [111] direction of aluminum and [0001] direction of sapphire, we convert Al unit cell from rock-salt structure to hexagonal structure while keeping the space group unaltered. The lattice constants of L, C, and R in the transverse direction are manually set as 4.8076Å, the optimal lattice constant of bulk sapphire. We want to highlight that we used force constants of Al/sapphire interface directly obtained from DFT calculation instead of the common simplified practice of using force constants of one material and only changing masses for the other. More details are included in the SI.

## ASSOCIATED CONTENT

**Supporting Information**

The supporting information includes the schematic diagram of the crystal structure of Al and sapphire, the AFM scanning image of the surface of Sub100, EBSD image of Sub200 showing only (111) oriented Al, the in-plane orientations of the Al films showing the presence of twinning, triple-axis diffraction results of Sub100 showing that Al film in Sub 100 (and Sub 200) is strain-free, and calculation details of the AGF modelling.

## AUTHOR INFORMATION

**Corresponding author**

*E-mail: sgraham@gatech.edu

**Notes**

The authors claim no completing financial interest.

## ACKNOWLEDGEMENTS

The authors would like to acknowledge the financial support from Office of Naval Research MURI Grant No. N00014-18-1-2429. Z.T. and R. H. would like to acknowledge the support by Office of Naval Research under ONR award number N00014-18-1-2724 and the Extreme Science and Engineering Discovery Environment (XSEDE) which is supported by National Science Foundation grant number ACI-1053575.

# Table of Content

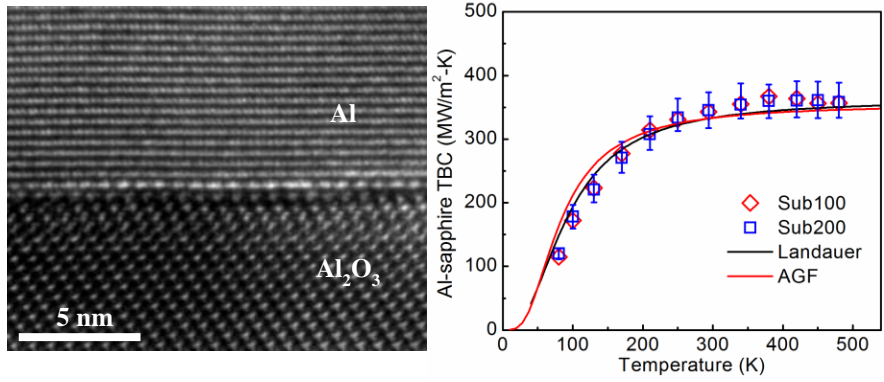